\documentstyle[11pt,epsf]{article}

\begin{document}

\title{Searches For Primeval Galaxies}

\author{C.A. Collins}

\maketitle

\footnotetext[1]{{\it Author's address:} Astrophysics Research Institute,
School of Engineering, Liverpool John Moores University, Byrom Street,
Liverpool L3 3AF, UK.}

{\it A primeval galaxy represents the earliest stages of a galaxy's life
and as such provides clues to the early history of the Universe and the
evolution of stars and galaxies. Over the last 20 years astronomers have
been engaged in the quest to detect the faint signals from these objects,
believed to lie at a distance comparable with the size of the Universe. A
wide variety of observational techniques have been employed in this search,
with astronomers eagerly awaiting each new generation of astronomical 
telescope or detector in the hope of finally solving the mystery to the
origin of galaxies -- or at least placing new and interesting constraints.
Until recently, primeval galaxies have eluded detection in these searches, 
however experiments over the last couple of years which use either 10m-class 
optical telescopes or state-of-the-art submillimetre array detectors, may 
hold the 
clue to the origin of structure as they have finally uncovered what appears 
to be a widespread population of young galaxies.}

\section{Introduction}


Primeval galaxies (hereafter PGs) can loosely be described as the progenitors 
of present-day galaxies, such as our own Milky Way, in the process of
assembling their mass through gravitational collapse and 
forming their first generation of stars. Detecting the starlight from
galaxies in such an early stage of their evolution has become one of 
cosmology's most elusive holy grails. The reason for this interest is that 
information about 
the ancestral 
characteristics of both spiral and elliptical galaxies helps
observational cosmologists in their quest to map out 
the star formation history of the Universe, which, in turn, provides
important constraints on galaxy formation theories. 

There are already excellent reviews of galaxy formation, galaxy evolution
and PG Searches in 
the astronomical literature:  Koo and Kron (1992), Pritchet (1994) -- on
which section 4 of this article is based, Sandage (1995) -- who provides an 
excellent historical account and Ellis (1997). The first part of this paper 
outlines the cosmological framework (section 2) and the physical processes 
underlying the growth of structure in the early Universe (section 3). The 
expected properties of PGs 
are described in section 4 and a summary of the previous
observational quests is then given (section 5). Recent insights into 
star-forming galaxies from deep optical surveys are discussed in
section 6 and the possibility of dusty PGs is reviewed as a solution to their
non-detection at other wavelengths (section 7). The paper
concludes with a discussion of the likely productive avenues of the future 
(section 8).

\section{Cosmological Distance and Time}

Searches for PGs are 
really attempts to detect the furthest recognisable galaxies and thereby
determine the epoch of galaxy formation. Let us begin by examining the 
relationship between the cosmological distance to a galaxy and the
light-travel time over that distance; this will illustrate the general power 
of PG searches to probe the early Universe. Astronomers measure the 
distance to a galaxy using the property known as the redshift ($z$). This 
is the term
given to the increase in wavelength of light as it propagates through 
space, caused by the expansion of the Universe. The size of the redshift is 
a measure of how fast
distant objects are moving away from the Earth and for distances not
large compared to the size of the Universe ($z<0.1$), the redshift can be 
related to the distance ($D$) by Hubble's law, $D \simeq cz/{\rm H}_0$, where $c$ is
the speed of light, and ${\rm H}_0$ is Hubble's constant. At larger
redshifts the look-back time to a galaxy starts to become significant 
compared to the age of the Universe and the dynamics of the expansion, 
determined by its energy density content, becomes progressively 
more important. 
The basic relationship between look-back time $\Delta t$ from
the present to a galaxy at redshift $z$ is given by
\begin{equation}
\Delta t = {\rm H}_0^{-1} \int^{z}_{0} (1+z)^{-1}[(1+z)^2(1+\Omega_{\rm
M}z) - z(2+z)\Omega_{\Lambda}]^{-1/2} dz,
\end{equation}
where $\Omega_{\rm M}$ and $\Omega_{\Lambda}$ are the contributions to
the energy density from the matter and cosmological constant respectively.
Figure 1 shows $\Delta t$ as a function of $z$ for three 
cosmological models which span the range of accecptable values for 
$\Omega_{\rm M}$ and $\Omega_{\Lambda}$. 
Despite the continuing debate in cosmology regarding the precise value 
of the quantities $\Omega_{\rm M},\, \Omega_{\Lambda}$ and ${\rm H}_0$,
observations of galaxies at $z\simeq1-5$ always probe back to a time when the 
Universe was $\simeq50\%-10\%$ of its 
current age respectively, in any plausible cosmology. This makes observations 
of galaxies at $z>1$ a very powerful tool for understanding the
conditions in the early Universe and, not surprisingly, observations of
distant objects, including searches for PGs, have over the 
years been extensive and wide-ranging: most, if not all, of the world's 
principal ground-based and space-born astronomical facilities, have, at some 
stage, devoted significant amounts of telescope time to primeval galaxy 
search programmes, covering the electromagnetic spectrum from the 
optical to the radio.

\begin{figure}[t]
\begin{picture}(100,270)
\put(60,260){\includegraphics{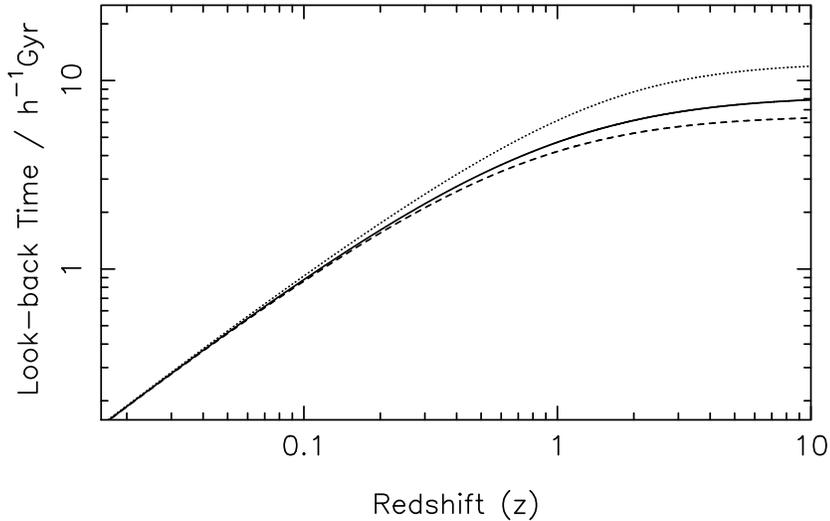}}
\end{picture}
\caption{Look-back time as a function of redshift for the cosmological models:
$\Omega_{\rm M}=1.0, \Omega_{\Lambda}=0$ (dashed line); 
$\Omega_{\rm M}=0.2, \Omega_{\Lambda}=0$ (solid line); 
$\Omega_{\rm M}=0.1, \Omega_{\Lambda}=0.9$ (dotted line). These 
models have been calculated for ${\rm H}_0=100h\,{\rm km}\,{\rm s}^{-1}$
Mpc$^{-1}$, where 
$h$ expresses the uncertainty in the value of ${\rm H}_0$ taking
values $h=0.5-1$ and $1\, {\rm
pc}\simeq3.1\times10^{16}\,{\rm m}$.}
\end{figure}


\section {Growth of Structure in the Big Bang Cosmology}

\subsection{The Cosmic Microwave Background}

The cosmic microwave background, discovered by Penzias and Wilson (1965), 
provides the strongest evidence for the hot origin of 
the Universe -- otherwise known as the hot Big Bang. In this grand model, the 
radiation and baryonic matter were in thermodynamic equilibrium at 
sufficiently 
early times, which involved the constituent fundamental particles 
interacting with photons via Compton 
scattering. As the Universe expanded and photons redshifted so the 
temperature of the background radiation cooled. The radiation field 
last interacted with matter at a redshift $z\simeq1000$, or some $10^6$ yrs 
after the Big Bang -- at which time the temperature of the radiation had 
cooled to $\leq 3000^{\circ}$ K, enabling electrons to be captured
forming neutral atoms. 

It is believed that structures such as
galaxies started out as small density fluctuations in this  
primordial soup of matter and radiation, 
growing by gravitational instability into larger overdensities as 
gravitationally bound systems were formed. The dense clumps of 
material in the 
gravitational potential wells caused collisional heating of the baryonic 
material, allowing rapid 
cooling of the gas by line radiation. Subsequently, larger clumps were formed 
as sub-clumps of material merged and combined -- with the final state 
of the bound material governed by angular momentum conservation. The first 
stars formed in the dense cores, which enriched the primordial gas with heavy 
elements as a result of supernovae explosions.

In recent years there has been substantial empirical underpinning of this 
general picture of the early growth of structure. For example, the earliest 
density perturbations have now been detected as small temperature 
fluctuations ($\delta{\rm T}/{\rm T} \simeq 10^{-6}$) in the cosmic
microwave background by the 
Cosmic Microwave Background Explorer (COBE) satellite (see Smoot and Keay 1993). In addition, semi-analytical 
models of the growth of 
structure, which incorporate most of the known physical processes, have had 
success 
in reproducing many of the optical properties of galaxies in the most distant 
surveys ({\it e.g.} Baugh {\it et al}. 1998).

\subsection{Density Inhomogeneities in the Early Universe}

The density fluctuations inferred from COBE are remarkably small and we know 
that for the overdensities in our own galaxy $\delta \rho/\rho>1$; 
furthermore, gravitationally collapsed structures, such as quasi stellar
objects, are known 
to exist at $z\simeq5$, so by this time 
at least some 
perturbations must have grown past $\delta \rho/\rho=1$. The big questions 
for galaxy formation then are: What is the scale-size of the first 
collapsing overdensities? How fast can fluctuations grow in the early 
Universe? What sort of fluctuations must they be?

In 1902 James Jeans considered the criterion for gravitational collapse in 
an infinite uniform medium in which small density perturbations ($d \rho $) 
give rise to adiabatic pressure changes or acoustic waves ($d p$), such  
that
\begin{equation}
d p = {\rm V}^2_{\rm s} d \rho,
\end{equation}

where ${\rm V}^2_{\rm s}$ is the adiabatic sound speed. The criterion for 
collapse is satisfied by a mass (known as the Jeans Mass, ${\rm M}_{\rm J}$) 
just large enough  
that the sound crossing time is larger than the free-fall collapse time
-- thus rendering the overdensity unable to respond with pressure changes 
fast enough to halt  
gravitational collapse. 


It is straight forward to apply the basic idea of 
Jeans to the problem of galaxy formation in the early Universe and 
calculate the minimum spherical mass which would stop expanding with the 
universal Hubble flow, turn around, and gravitationally collapse. At
$z=1000$, just after decoupling 
\begin{equation}
{\rm M}_{\rm J} \simeq 2\times 10^4 (\Omega_0 h^2)^{-1/2}\,{\rm
M}_{\odot}, 
\label{eqn:jeans2}
\end{equation}

where $\Omega_0$ is the present mass density, and ${\rm M}$ is the unit of 
solar mass ($1\, {\rm M}_{\odot}=2\times10^{30}\, {\rm kg}$). 

This value assumes that 
the energy density of the Universe is matter dominated -- at epochs earlier 
than about the decoupling stage ($z>1000$ or $t\leq10^6$ yrs), the energy density 
of the radiation field exceeds 
that of the matter content and the sound speed is then relativistic.
During this period, known as the radiation dominated era, the minimum mass 
required to undergo 
gravitational collapse of baryonic overdensities rises to
\begin{equation}
{\rm M}_{\rm J}\simeq10^{14} (\Omega_0 h^2)^{-2}\, {\rm M}_{\odot}. 
\label{eqn:relmas}
\end{equation}

This mass is so large that it is in fact comparable to the entire baryonic
content of each causally connected volume of space at any time during the 
radiation 
dominated era.  Furthermore, the characteristic timescale required for 
matter fluctuations to collapse is longer than
the characteristic time for the expansion of the Universe during this era. 
A combination of the large Jeans mass and rapid expansion
time of the Universe guarantee that no baryonic fluctuations 
gravitationally
collapse for the first $10^6$ yrs after the Big Bang and galaxy formation 
really only begins after this time at the onset of the matter dominated 
era. 

%

An estimate of the redshift at which galaxies collapse can be made using
the Jeans criterion: an overdensity larger 
than ${\rm M}_{\rm J}$ will stop expanding and collapse in a time 
$t_c$, given by 
\begin{equation}
t_c = \left ( \frac{3 \pi}{32 {\rm G} \rho} \right )^{1/2}. \label{eqn:coltime}
\end{equation}

Taking the example of our own Milky Way, the total mass implied
from dynamical studies is $5\times10^{11}\,{\rm M}_{\odot}$ and from the 
distance
of the furthest stars in the galactic halo a plausible size to the Galaxy 
before it collapsed is $\simeq 100$ kpc. This implies a density
$3.6\times10^{-24}\,{\rm kg}\,{\rm m}^{-3}$ and, using equation
(\ref{eqn:coltime}) a collapse time 
of $\simeq10^9$ yrs, which corresponds to a look-back time of 7.2 Gyr if 
$\Omega_{\rm M}=
0.2$ and 5.4 Gyr if $\Omega_{\rm M}=1.0$ (all assuming
$\Omega_{\Lambda}=0, h=1$). From figure 1 this
translates to redshifts of formation $z_f\simeq 4-2$ respectively. If
$h=0.5$ then formation redshifts range from $z_f\simeq 6-4$ for the same 
mass densities. Finally, bound objects the size of galaxies are unlikely to 
exist much above these redshifts for the simple reason that the age of the 
Universe is then short compared to the free-fall collapse time.

\subsection{Non-Baryonic Dark Matter}

It will not have escaped the attention of astute readers that the above
physical arguments make no mention of the well known controversy regarding 
the possibility of missing mass in the Universe and 
the type of matter which might dominate the overall mass density. The most 
likely form
of non-baryonic dark matter to dominate the mass density is known as cold dark 
matter, the primary candidates for which are in the form of Weakly 
Interacting Massive
ParticleS or {\rm WIMPS} -- such as the super-symmetric lepton partners 
of bosons ({\it e.g.} Photino, Gravitino or Higgsino), which have predicted 
masses $\sim{\rm GeV}$. If the
major constituent of $\Omega_{\rm M}$ in the Universe is in the form of
cold dark matter then the growth of 
fluctuations begins before decoupling: the non-baryonic matter 
decouples from the radiation earlier and because it feels no radiation pressure can get on 
with the task of collapsing and forming gravitational potential wells during 
the radiation dominated era. At the onset of decoupling the baryonic gas,
which until this time has been locked to the radiation field by 
electrostatic forces, falls very quickly into the gravitational potential 
wells created by the non-baryonic matter. In this case, assuming that the 
amplitude of density fluctuations decreases with scale, the 
first structures to form have a mass $\simeq 10^5\,{\rm M}_{\odot}$ --
similar to the baryon-dominated case and again set by the 
pressure of the baryons after recombination. 


To summarize, the variety of models represented in this 
short discussion of the growth of structure suggests that galaxies
form by the collapse and subsequent merging of sub-galactic size units 
over a range of formation redshifts ($ 2 \leq z_f \leq6$). Uncovering PGs is 
one of the few ways of providing constraints on the various
possibilities.

\section{Properties of PGs}

The two fundamental requirements of any PG search are a sound
search strategy -- covering a sufficient area of sky to a sufficient depth, 
and an ability to recognise a PG amongst the background of 
older and more normal galaxies. Therefore, it will be useful to have
some idea of what the basic properties of such and object might be.

\subsection{Surface Density}

An estimate of the surface density of PGs on the sky can be made by 
extrapolating the volume density of local galaxies 
($\sim 0.015\,h^3{\rm Mpc}^{-3}$) to high redshift. Suppose that all galaxies 
form at a particular redshift by gravitationally collapsing 
over a period of 1 Gyr, rapidly forming their first generation of stars: 
then the resulting areal density of PGs as a function of their formation 
redshift is shown in figure 2. PGs should be very abundant irrespective of the
uncertainties in the cosmology, with at least $10^3$ in each area of sky
the size of the moon!

\begin{figure}[t]
\begin{picture}(100,270)
\put(60,260){\includegraphics{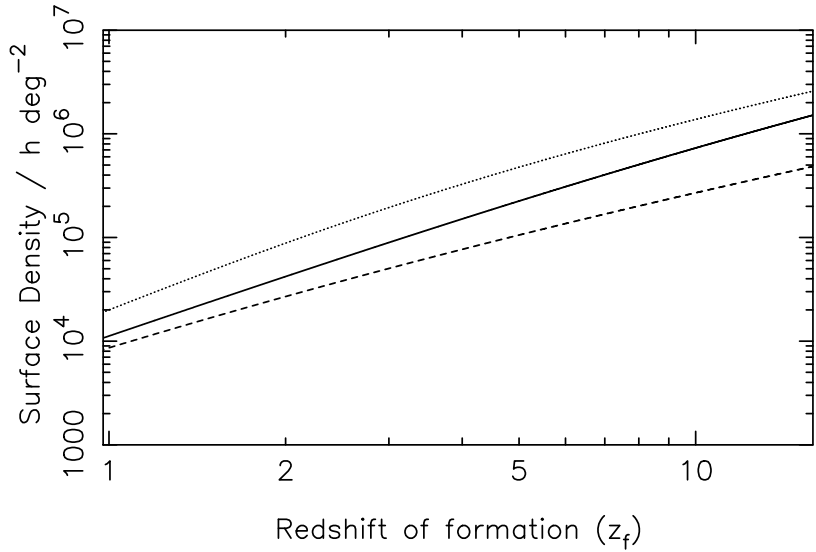}}
\end{picture}
\caption{Expected surface density of PGs as a function of their formation
redshift, assuming a collapse time of 1 Gyr. The three curves represent
the same cosmologies as in Figure 1.}
\end{figure}

\subsection{Angular Size}

This is one of the most difficult properties to estimate. The angular size
of a PG depends upon when the bright phase occurred relative to the
collapse as well as the redshift and cosmology. If the star 
formation occurs soon after the onset of collapse then PGs could be
low-surface brightness objects with angular diameters $10-30$ arcsec
(Partridge and Peebles 1967). At $z\geq5$ and with a surface density of $\geq10^5\,{\rm
deg}^{-2}$ (see figure 2) PGs would be virtually overlapping on the sky
and very difficult to detect as discrete sources using conventional
techniques. Possibly the best way
to to detect such objects is to investigate the integrated background
light resulting from such a population of sources and experiments to do
just this have been carried out. At the other extreme is the
idea, stretching back a number of years, that the star formation rate
(hereafter SFR) is closely coupled to the gas density and hence the bulk of 
the star
formation occurs in the central regions of collapsed PGs (Eggen,
Lynden-Bell and Sandage 1962, Larson 1974). In this case we can expect the 
visible signature of PGs to have an angular diameter $\simeq0.1-0.5$
arcsec, making them very difficult to resolve from the ground due to the 
limitation imposed by atmospheric ``seeing'' irregularities. Such objects
would be excellent targets for optical satellite missions, such as the
Hubble Space Telescope with its diffraction-limited
0.1 arcsec imaging capability.

\subsection{Luminosity}

The principal energy source of PGs is the release of binding energy in
nuclear reactions at the centres of hot young stars. A crude estimate of
the luminosity of a young galaxy can be obtained by simply considering the
energy required to convert a fraction $Z$ of hydrogen into metals, based
on the metalicities of old stars in the Milky Way, for which 
$Z\simeq0.01$. A 
PG of mass $10^{11}{\rm M}_\odot$, with a star-forming phase
lasting 1 Gyr (equivalent to a SFR of about 100 M$_{\odot}$
yr$^{-1}$) has a net bolometric luminosity $\simeq 6\times10^{37}{\rm W}$. This makes PGs 10-100 times brighter than
nearby bright spiral galaxies in which stars are being formed at a rate
of a few ${\rm M}_\odot$ yr$^{-1}$. At $z=5$ this corresponds to an
optical ($4500$ \AA) flux of $10^{-18}\,{\rm W m}^{-2}$ -- at least
100 times fainter than the terrestrial night-sky background at 
these wavelengths; while in the near-infrared ($1-5\,\mu$m) the Earth's
atmosphere is $\sim1000$ times brighter than at optical wavelengths, 
making PGs at least $10^5$ fainter than the sky! For these reasons PGs are 
unlikely to be masquerading in existing catalogues of bright galaxies and can 
only hope to be detected by the largest telescopes.

\subsection{Spectral Energy Distribution}

In recent years significant progress has been made in reliably modelling
the evolution of stellar populations in galaxies ({\em e.g.} Charlot {\it et
al}. 1996). The basic approach, known as ``isochrone synthesis'',
involves computing the evolutionary tracks of stellar populations for an
instantaneous star burst ({\it i.e.} with no age dispersion) incorporating 
a finite rate of star formation. In this way the distribution of stars of 
various masses and ages can be modelled smoothly with time and isochrone
synthesis models are found to reproduce well the observations of stellar
populations in nearby galaxies if they formed more than a few Gyrs ago.
This technique has been widely used to predict the major spectral
characteristics of young galaxies $\leq1$ Gyr after formation. Figure 3
shows the resultant spectral energy distributions for a galaxy forming
stars at a rate of $1\,{\rm M}_{\odot}\,{\rm yr}^{-1}$ observed after
ellapsed periods ranging from $4\times10^7$ yrs to $10^{10}$ yrs. The 
essential characteristic of figure 3 is that genuinely young
galaxies radiate an almost constant energy density from $0.1-2.0\mu{\rm
m}$. The rise in the relative emission at longer wavelengths as the
stellar population ages is due to the
death of massive hot stars, whose lifetime ($t$) varies with mass (M) as
$t\simeq({\rm M}/{\rm M}_{\odot})^{-2}\times10^{10}$ yrs, which evolve into 
cooler red giant stars and supernovae. One important consequence of the flat 
spectral energy distribution
is that the measured optical or near-infrared flux of a PG is a direct measure of
the instantaneous SFR within the system. 

\begin{figure}[t]
\begin{picture}(100,270)
\put(60,260){\includegraphics{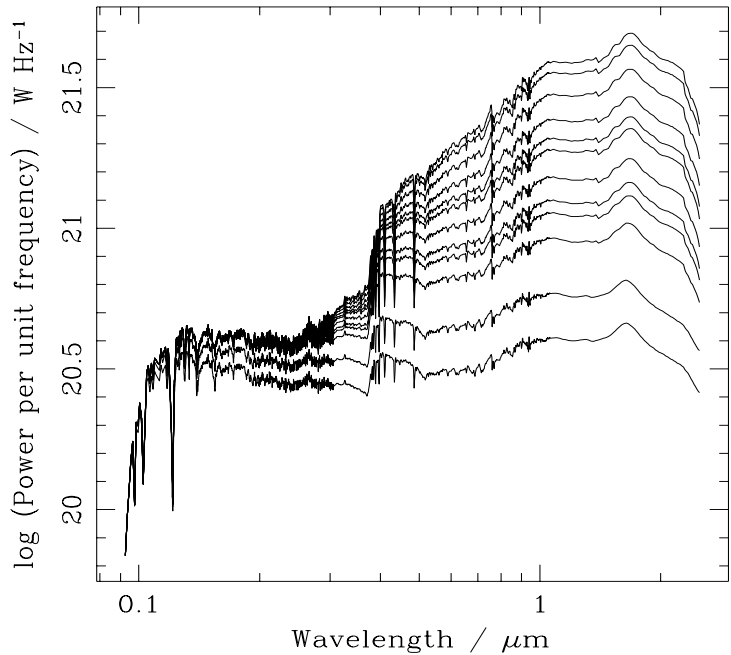}}
\end{picture}
\caption{The spectral energy distribution expressed as power per unit
frequency interval computed from stellar
population synthesis code (see text) for a galaxy with mass
$10^{11}\,{\rm M}_{\odot}$ and a constant SFR of $1\,{\rm
M}_{\odot}$yr$^{-1}$ observed after a time (yrs): $4\times10^7$ (bottom
curve), $10^8$, $3\times10^8$, $5\times10^8$,
$6\times10^8$, $10^9$, $2\times10^9$, $3\times10^9$, 
$5\times10^9$, $8\times10^9$, $10^{10}$ (top curve).
In these models Lyman $\alpha$ at $1215\,{\rm \AA}$ is shown in absorption. No
extinction due to dust has been assumed (see figure 5)}
\end{figure}

In addition to the stellar continuum emission, young star forming regions
within the Milky Way and other nearby galaxies exhibit intense line
radiation, principally the hydrogen recombination lines of Lyman $\alpha$ at 
$1215\,{\rm \AA}$ and the Balmer line H$\alpha$ at 6562 \AA. The Lyman $\alpha$
line is shown in absorption in figure 3, but in star-forming regions these 
lines are seen in emission associated with the ionised hydrogen in the 
surrounding gas which constitutes 
the star-forming nebulae. The intensity of these lines is a measure of the 
ambient ionizing UV flux from young hot stars and can also be used as a 
tracer of star formation to estimate the SFR in an
independent way from the spectral energy distribution (Kennicutt {\it et
al}. 1987).

In conclusion, although predictions are sketchy, there are certain  
characteristics which PGs are likely to have that one can highlight: PGs should
not be rare objects and can be identified in the optical or near-infrared by their 
flat continuum emission or intense line radiation. On the other hand,
despite being intrinsically luminous objects, at least compared to
normal galaxies, the likely formation redshift of PGs 
dictates that they are almost certainly faint and possibly of low-surface 
brightness, making them very difficult to detect.

\section{Searches For PGs -- An Historical Overview}

It would appear from the previous discussion concerning the likely
properties of PGs that a search covering only a limited area
of sky, but reaching to faint flux limits, most closely represents the optimum 
search strategy. The high number density of PGs expected also ensures
that one part of the sky will be pretty much as good as any other in
terms of the likelihood of detection -- although clearly avoiding regions
of high galactic extinction and high stellar density, such as the plane
of the Milky Way or towards the Large Magellanic Clouds, is advisable. PG 
searches have been carried out for more than 20 years
and have always used the largest, and therefore most sensitive, telescopes
and accompanying instruments to conduct such surveys. The observations
have always been difficult and the story is one of
astronomers, goal-oriented in their quest to uncover the earliest
population of galaxies, only making significant observational inroads
when accompanying technological advances in telescope or instrument design 
take place. The first serious searches for PGs, carried out in 
the 1970s, utilised the large areal coverage obtainable with photographic 
plates (Partridge 1974), although it was only when optical charge-coupled
device (CCD) technology was 
developed, in the early 1980s, that really useful limits were first reached.

\subsection{Optical continuum and emission-line searches}

A number of experiemnts have been carried out designed to detect the
emission from redshifted Lyman $\alpha$ -- possibly the strongest
single emission feature from young galaxies. One group of these observations 
utilises the technique of narrow-band imaging: a CCD image taken with a filter centred on redshifted Lyman $\alpha$ with 
a passband well-matched to the expected width of the emission line ($\simeq
20$ \AA), greatly enhances the contrast of the line against 
the continuum of the galaxy and, more importantly, against the terrestrial 
sky-emission. In this way Pritchet and Hartwick (1987) set strong limits on 
the space density of PGs at $4.5<z<6$ from narrow-band imaging at wavelengths
$6000\leq\lambda\leq8000\,{\rm \AA}$ using the Canada-France-Hawaii 3.6m
telescope in Hawaii, while results on the same telescope by De Propris
{\it et al.} (1993) using a system of narrow-band filters covering 
$4000\leq\lambda\leq5000\,{\rm \AA}$ found upper limits on
Lyman $\alpha$ emission that are inconsistent with the simple model predictions
discussed above by a factor of about 10 at redshifts $2\leq z \leq3$.

The other technique designed to detect redshifted 
Lyman $\alpha$ from PGs utilises a grating spectrograph incorporating a long 
object slit to maximise sky coverage during the observation. While the sky 
coverage obtainable is significantly less than that of the narrow-band 
technique, the advantages are that a significantly 
larger wavelength range can be observed simultaneously and the sky
background can be subtracted accurately. For example, Thompson and Djorgovski 
(1995) used spectroscopic data from the 200 inch Hale telescope at the
Palomar Observatory, obtained for other purposes over the
course of a 7-year period, to search for serendipitous Lyman $\alpha$
emission over the wavelength range 
$5000\leq\lambda\leq 7500\,{\rm \AA}$ and covering a total area of 15 arcmin$^2$ of
sky (sufficient to contain at
least 20 PGs) -- placing limits on the SFR in
PGs of $100\, {\rm M_{\odot}}$yr$^{-1}$. 

Accompanying experimental limits on the brightness of the extragalactic
background light at optical wavelengths are also consistent with the
lack of success in detecting individual PGs at optical wavebands, helping
to rule out the existence of a substantial PG population at $z<5$ ({\it
e.g. } Dube, Wickes and Wilkinson 1977).

\subsection{The Era of the Near-Infrared}

Working at near-infrared wavelengths provides the opportunity
to probe to even higher redshifts; the optical searches being limited to
finding objects at $z\leq5$. The earliest results in the near-infrared 
($1-5\,\mu{\rm m}$) were limited to 
experiments with single-element detectors: Boughn, Saulson and Uson (1986) 
placed some of the earliest limits at $2\mu$m by searching for  
the fluctuations in the sky-brightness on scales $10-30$ arcsec -- an
experiment optimised to detect the signal of low surface brightness 
PGs at $z\sim20$ overlapping on the sky. Collins and Joseph (1988) carried
out what was probably the first near-infrared search for discrete PGs,
looking for candidate objects with the expected flat spectral energy distribution
(see figure 3).
A summary of these early results is presented by Thompson, Djorgovski and 
Beckwith (1994), who conclude that with the limited field size and
sensitivity available to
near-infrared instruments at that time, the expected SFR of PGs 
could still be as high as $1000\,{\rm M}_{\odot}$yr$^{-1}$. 

An important technological landmark for PG searches came with the
advent of CCD-type array cameras for use in the  
$1-5\,\mu{\rm m}$ waveband; in 1986 the 3.8m United Kingdom Infrared Telescope
(UKIRT) in Hawaii was the first telescope to have such a general-purpose
camera. In fact infrared array devices were constructed as early as 1974
but the technical information came late to astronomers due to the 
secrecy surrounding the early development which was largely carried out
for military applications. 

With these new panoramic detectors near-infrared searches began to rival their
optical counterparts in limiting sensitivity and areal coverage: Parkes,
Collins and Joseph (1994) placed the first really useful limits on the
number density and Lyman $\alpha$ luminosity of objects at $7\leq z \leq9$
using narrow-band filters, while Pahre and Djorgovski (1995) achieved
similar flux limits by targetting the emission lines of
H$\alpha$, [OII], H$\beta$ and [O II] redshifted to $z=2.88$ and
$z=4.79$ using the Keck 10m telescope situated alongside UKIRT in Hawaii. 
These limits constrain the 
SFR of PGs to a value $\sim1-10\,{\rm M}_{\odot}$yr$^{-1}$, similar to
the constraints achieved in the optical.

One final avenue of investigation previously undertaken concerns radio
searches for neutral hydrogen: 
From the lack of absorption lines at wavelengths shorter than Lyman $\alpha$ 
in the spectra of high
redshift quasi stellar objects, it has been known for some time that the mass density in
diffuse neutral hydrogen must be very small (Gunn and Peterson 1965).
Nonetheless, a number of experiments have been carried out designed to detect 
redshifted 21cm radiation from clumps of neutral hydrogen corresponding to 
an evolutionary stage of PGs preceeding the epoch at which 
stars form (Hogan and Rees 1979; Davies, Pedler and Mirabel
1978; Uson, Bagri and Cornwell 1991). A big drawback of these
experiments as they currently stand is limiting sensitivity -- the
faintest of which corresponds to a neutral hydrogen mass $\sim10^{14}{\rm
M}_{\odot}$ at $z\geq3$, which is
much bigger than any predicted characteristic mass scale (see section 3).

\subsection{Masquerading PGs}

Before looking at what a null
detection of the starlight from PGs might mean it is worthwhile to consider
if the problem lies in misunderstanding the nature of PGs and whether it
is possible that PGs are masquerading as extragalactic 
objects of a type not yet considered: some galaxies exhibit poweful radio 
emission, resulting from a
relativistic jet eminating from the core of the galaxy in which a black
hole is believed to reside. This radio emission can be detected out to 
great distances and some of these objects are believed to be the site of 
large amounts star-formation: the source 3C326.1 at $z=1.8$ is forming stars 
at the rate of $300\,{\rm M}_{\odot}$ yr$^{-1}$ (McCarthy {\it et al}. 1987); while 
the source called 4C41.17 at $z=3.8$ emits $10^{13}$ L$_{\odot}$ 
of starlight (where $1 {\rm L}_{\odot}\simeq3.8\times10^{26}$ W) and contains enormous amonts of dust 
(Dunlop {\it et al}. 1994). Even though these objects are a powerful probe of
the general characteristics of galaxies at high redshift, their strong radio 
signature isolates them as a special class of galaxy -- too extreme to be
the ancestors of more normal spirals and ellipticals.

Quasi stellar objects (QSOs) are at least candidates on the grounds
that they occupy a similar redshift range to that expected of PGs -- the
furthest QSO has a redshift close to 5 -- however, the comparison cannot be 
taken further than that; QSOs are powered by material accretion onto a black 
hole and
their optical/near-infrared emission is dominated by the output from this
process rather than starlight. What QSOs do tell us is that the high
abundance of heavy elements, inferred from the absorption lines in their
spectra, indicates significant chemical enrichment of the Universe along
random lines-of-sight before $z\simeq4$. The lack of a strong resemblance 
between other high-redshift
objects such as QSOs and the expected characteristics of PGs forces us to 
conclude that PGs are not being mistaken for other
extragalactic sources in large numbers and it is necessary to look for other 
reasons for their non-detection.

\section{Evidence for Type-Dependent Evolution of Galaxies}

\subsection{The Lyman $\alpha$ dropouts}

Although PG searches over the last 20 years have proved
disappointing in uncovering a distant population of monolithic PGs, it is
precisely this lack of success which has given rise in recent years to 
renewed interest in galaxy evolution at more modest 
redshift ($z\simeq1-3$), where galaxies are detected in numerous 
quantities by, for example, the Hubble Space Telescope. 

These studies of evolution have 
proved pivital in our understanding 
the evolution of local galaxies, not by way of uncovering PGs as such --
this is one discovery which has alluded the Hubble Telescope so far,  
but by measuring the rate at which stars are formed in normal spiral galaxies  
over a look-back time of a few Gyrs and have produced a rapid advance in
studies of the Universe in the era of the 10m-class telescopes.

The basic technique used to find galaxies at these redshifts 
involves broad-band optical imaging in a number of filters to detect the
sudden drop in the spectral energy distribution shortward of the
redshifted Lyman break ($\leq 900(1+z)\,{\rm \AA}$). Photons with shorter
wavelengths than this ionize hydrogen and are re-processed primarily into Lyman
$\alpha$ photons. The resultant characteristic spectral ``step'' can be seen 
in figure 3 for a zero redshift galaxy. In a
series of experiments starting in 1992, Steidel and collaborators (see
Steidel {\em et al}. 1996) have used this technique along with follow-up
deep spectroscopy on the Keck telescope, to confirm the existence of a
substantial population of compact ($\sim0.4\,{\rm arcsec}$) star-forming galaxies at $z\sim3$. From the strength of their emission lines and shape of
their continua, these galaxies have SFRs $\leq 30\,{\rm
M}_{\odot}$yr$^{-1}$ and a number density equivalent to between $50\%-10\%$ 
of the space density
of present day bright galaxies. Progress in this area has been very rapid and 
it is now possible for the first time to sketch out the star-forming
history of a substantial fraction of the present day galaxy population
(Madau {\it et al}. 1996): The work of Steidel, together with other samples selected in a
similar manner -- including
galaxies found by the Hubble Space Telescope (e.g. Lilly {\it et al}. 1996, 
Connolly {\it et al}. 1997), indicate that the overall SFR of galaxies
increases from $z=0$ to $z=2$, during which time a significant fraction
of the heavy elements in the Universe are formed, and then tails off
towards higher redshift -- see figure 4 taken from Madau {\it et al.} (1998). 
Although the uncertainties are considerable and many surveys are still 
underway, figure 4 is a real landmark in cosmology as it represents the very 
first attempt to map out the star-forming history of galaxies.

\begin{figure}[t]
\begin{picture}(100,270)
\put(60,260){\includegraphics{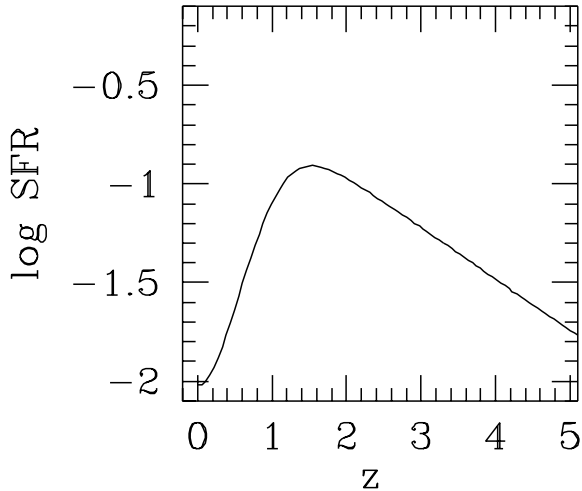}}
\end{picture}
\caption{The implied star-formation rate (SFR) in M$_{\odot}$yr$^{-1}$ against
redshift inferred from several optically selected galaxy samples. The
plot is taken from Madau {\it et al.} 1998.}
\end{figure}


The Lyman dropout galaxies are known to be young from the shape of
their spectral energy distributions, which are quite flat (see figure 3), 
and they also show signs of energetic outflows consistent with intense
bursts of star formation (Pettini {\it et al.} 1998). So in key respects this
widespread population of objects at $z\sim3$ has some of the essential
characteristics which we expect from PGs and it is quite
possible that we are seeing directly, for the first time, the formation
of present day galaxies. These results certainly demonstrate beyond any
reasonable doubt that massive galaxy formation was well underway by
$z=3.5$. This relatively late epoch of galaxy formation appears to be in general 
agreement with the popular models of galaxy formation in which the mass density 
of the Universe is dominated by cold dark matter, discussed in section 3, 
which predict a peak SFR of a few ${\rm M}_{\odot}$yr$^{-1}$ between 
$2 \leq z \leq 4$ (Baugh {\it et al}. 1998).

\subsection{The Age of Elliptical Galaxies}

Despite the discovery of the Lyman dropout population at moderate redshift, 
the alternative view that galaxies formed at high redshift in a short but 
intense period of star formation still provides the best explanation for the 
formation of at least some galaxies -- in particular, elliptical 
galaxies. These constitute $\sim 30\%$ of the galaxy population seen at low
redshift and ellipticals seen at $z\simeq0.5$ have spectral energy distributions 
conststent with a passively evolving stellar population with an age of 
several Gyrs (see figure 3), equivalent to a formation redshift $z\geq5$ ({\em e.g.} 
Bender {\it et al}. 1996, Ellis {\it et al}. 1997). So a complete picture of when 
galaxies form remains elusive: nearby elliptical
galaxies, the most massive of any galaxy-type, contain a stellar
population too old for them to be the low redshift counterparts of the
Lyman dropout galaxies and yet primeval elliptical galaxies are not seen in 
the deep optical or near-infrared searches for 
monolithic PGs. A key 
question then is: how can the dearth of PGs in the optical and 
near-infrared searches (described in section 5), many of which are 
tuned to detect objects at $z\simeq5$, be reconciled with the 
seemingly inevitable conclusion from evolution studies that this is exactly 
the epoch at which elliptical galaxies are expected to form?
 
\section{Dusty Elliptical PGs}

By far the most likely answer to this question is that young elliptical
galaxies are enshrouded in the dust formed by young stars, which absorbs
the starlight, making them almost invisible at optical and near-infrared
wavelengths. Figure 5 shows the effect of only a small amount of dust on 
the optical and UV light eminating from a young galaxy 1 Gyr old. The 
absorbed energy simply gets re-radiated as thermal
emission giving rise to substantial emission at far-infrared and
sub-millimetre wavelengths.

\begin{figure}[t]
\begin{picture}(100,270)
\put(60,260){\includegraphics{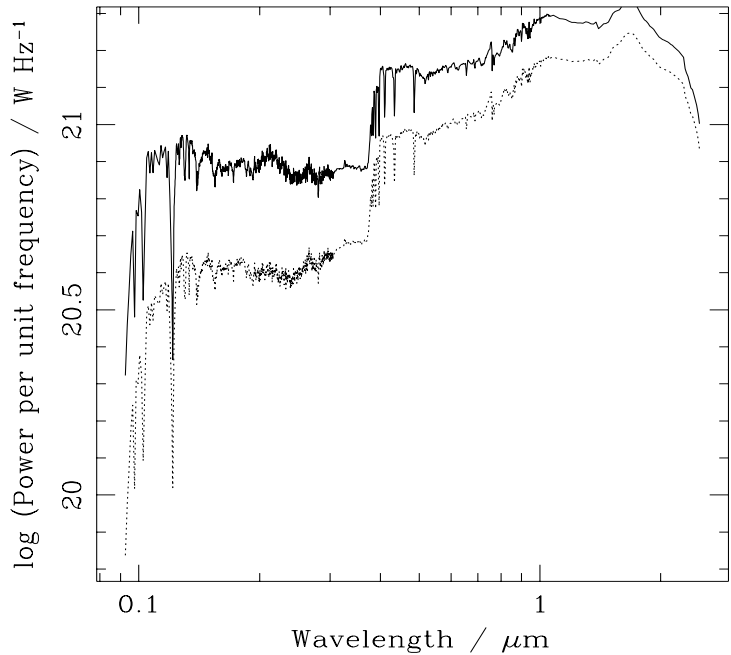}}
\end{picture}
\caption{The spectral energy distribution of a galaxy 1 Gyr old forming
stars at a rate of $1\, {\rm M}_{\odot}$yr$^{-1}$. The top curve is the
prediction with no dust (as in figure 3) and the bottom curve the prediction 
assuming a simple dust creation model.}
\end{figure}

In a parallel way to that of near-infrared astronomy some 10 years ago, recent 
technological developments in the last couple of years have led to the first 
searches for dusty PGs: SCUBA (the Submillimetre Common User Bolometer Array) 
was built by the Royal Observatory Edinburgh for use on the James Clerk Maxwell
Telescope (JCMT) in Hawaii. It is the world's largest submillimetre camera, 
and takes images
simultaneously at two submillimetre wavelengths. It can take deeper
images of the sub-millimetre ($450\,\mu$m$- 2.0$ mm) sky than ever before,
mapping areas in minutes which previously used to take hours to 
complete and is quite capable of detecting re-processed starlight from dusty
PGs at high redshift. 

Fueling the excitment surrounding the anticipated arrival of SCUBA was the 
report of the first detection of an extragalactic far-infrared background
from the COBE satellite (Puget {\it et al}. 1996). 
This background appears isotropic over the range $400\,\mu$m$- 1.0$ mm
and has a natural explanation in terms of re-radiated starlight from dusty
galaxies at high redshift (see Puget {\it et al}. 1996). Despite only 
begining regular observations in May
1997, SCUBA has not proved disappointing: detections of sources with 
submillimetre spectral
properties consistent with high redshift dusty PGs were reported last year 
(Smail {\it et al}. 1997). Some caution is necessary in the early
interpretion of these results since the Smail
{\it et al} search fields cover only a few arcmin$^2$ of sky and were not
the traditional blank fields characteristic of previous PG searches, but
instead were centered on known distant clusters of galaxies. This was
done in order to utilise the familiar gravitational lensing effect seen
in rich clusters, which amplifies the light from backgound objects as it
passes through the cluster's gravitational potential well. While this
makes distant galaxies brighter and therefore easier to detect, source
luminosities and surface density estimates require accurate knowledge of
the amplification factor, which is intrinsically uncertain. 

Very recently, the results from 
two independent blank field searches have been
reported (Hughes {\it et al}. 1998, Barger {\it et al.} 1998). Each group
surveyed approximately $10\,{\rm
arcmin}^2$ of sky (including the celebrated Hubble Deep Field). These data 
reveal objects with submillimetre spectral properties consistent 
with that expected from dusty galaxies in the
redshift range $z=2-4$, powered by star formation with an implied SFR
$5-10$ times larger than the Lyman dropout galaxies seen in optical
surveys. Results from surveys covering larger areas of sky are eagerly
awaited: if these and similar experiments confirm a dusty PG population, the 
last empirical ingredient required to understand the formation of galaxies 
may have been found and a description of galaxy 
evolution over a wide span of cosmic time would then be within reach.

\vspace*{0.1in}

\section{Concluding Remarks and Future Prospects}

This is a very rapidly changing field of research. The advent of the
10m-class optical telescopes such as Keck, soon to be joined by
others, has revolutionised our understanding of galaxy evolution in
recent years and much still can be done in the optical and near-infrared: 
the recent discoveries of Lyman $\alpha$ emission from a few galaxies at 
$z=3-6$ (Dey {\it et al}. 1998, Hu {\it et al}. 1998), shows that not all
galaxies at this epoch 
are dusty, which provides fresh hope for the discovery of PGs when the
new generation of even
larger near-IR arrays becomes available on the largest telescopes. Certainly
the US space agency NASA recognises the importance of continuing PG
searches: over the next 
two decades it has dedicated itself to the ``cosmic
origins'' programme, designed to answer such questions as ``how did the first 
galaxies form?''This will require a succession of sophisticated telescopes, 
each building on the results of previous missions augmented with ground-based 
observations. One such project is the New Generation Space Telescope --
a 4m (or possibly 8m) descendant of the incumbent Hubble Space Telescope, designed for
diffraction-limited imaging over $1\,{\rm deg}^2$. With such
capability it will be possible to observe the individual supernovae explosions
resulting from massive bursts of star formation happening at almost any 
epoch.

Undoubtedly, far-infrared and submillimetre observations will have future
r\^{o}le to play in the unfolding story: the Planck satellite mission (named
after the German physicist Max Planck), is
part of the European Space Agency's Horizon 2000
Scientific Programme. Due to be launched in $\simeq2005$, Planck will 
survey the whole sky at millimetre wavelengths with
unprecedented sensitivity and angular resolution. Designed primarilly 
to map fluctuations in the cosmic microwave background, Planck has the
capability to detect small fluctuations in the far-infrared background 
tentatively detected by COBE, thereby distinguishing between rival
theories predicting different epochs of
galaxy formation. Perhaps the most exciting prospect on the timescale of
10 yrs or so, is the development of 
millimetre array
imagers, capable of both high spatial and spectral resolution imaging
over the whole sub-millimetre waveband. For example, the National Radio 
Astronomical Observatories millimetre array, proposed to the US National
Science Foundation will provide a spatial resolution of 10 milliarcsec
between $10\,{\rm mm} - 350\,\mu$m and a sensitivity capable of detecting
the dust emission from a bright star-forming galaxy with luminosity 
$>10^{11}$L$_{\odot}$ to $z=20$.

\newpage

{\bf Acknowledgements}

I would like to thank Gordon Rogers for the spectral energy distribution
plots and Doug Burke for useful discussions.

{\bf References}

Barger, A.J., Cowie, L.L., Sanders, D.B., Fulton, E., Taniguchi, Y.,
Sato, Y., Kawara, K., and Okuda, H., 1998, {\it Nature}, {\bf 394}, 248.\\
Baugh, C.M., Cole, S., Frenk, C.S., and Lacey, C.G., 1998, {\it
Astrophysical Journal}, {\bf 498}, 504.\\
Bender, R., Ziegler, B., Bruzual, G., 1996, {\it Astrophysical Journal},
{\bf 463}, L51.\\
Boughn, S.P., Saulson, P.R., and Uson, J.M., 1986, {\it Astrophysical
Journal}, {\bf 301}, 17.\\
Charlot, S., Worthey, G., Bressan, A., 1996, {\em Astrophysical Journal},
{\bf 457}, 625.\\
Collins, C.A., and Joseph, R.D., 1988, {\em Mon. Not. R. astr. Soc.},
{\bf 235}, 209.\\
Connolly, A.J., Szalay, A.S., Dickinson, M.E., Subbarao, M.U., and
Brunner, R.J., 1997, {\em Astrophysical Journal}, {\bf 486}, 11.\\
Davies, R.D., Pedlar, A., and Mirabel, I.F., 1978, {\em Mon. Not. R.
astr. Soc.}, {\bf 182}, 727.\\
De Propris, R., Pritchet, C.J., Hartwick, F.D.A., and Hickson, P., 1993,
{\em Astronomical Journal}, {\bf 105}, 4.\\
Dey, A., Spinrad, H., Stern, D., Graham, J.R., and Chaffee, H., 1998,
{\it Astrophysical Journal}, {\bf 498}, L93.\\
Dube, R., Wickes, W.C., and Wilkinson, D.T., 1977, {\it Astrophysical
Journal}, {\bf 215}, L51.\\
Dunlop, J.S., Hughes, D.H., Rawlings, S., Eales, S., and Ward, M.J.,
1994, {\em Nature}, {\bf 370}, 347.\\
Eggen, O.J., Lynden-Bell, D., and Sandage, A.R., 1962, {\it Astrophysical
Journal}, {\bf 136}, 748.\\
Ellis, R.S., 1997, {\em Ann. Rev. Astron. Astrophys.}, 1997, {\bf 35}, 389.\\ 
Ellis, R.S., Smail, I., Dressler, A., Couch, W.J., Oemler, A., Butcher,
H., and Sharples, R.M., 1997, {\it Astrophysical Journal}, {\bf 483},
582.\\
Gunn, J.E., and Peterson, B.A., 1965, {\em Astrophysical Journal}, {\bf
142}, 1633.\\
Hogan, C.J., and Rees, M.J., 1979, {\em Mon. Not. R. astr. Soc.}, {\bf
188}, 791.\\
Hughes, D., Serjeant, S., Dunlop, J., Rowan-Robinson, M., Blain, A.,
Mann, R.G., Ivison, R., Peacock, J., Efstathiou, A., Gear, W., Oliver,
S., Lawrence, A., Longair, M., Goldschmidt, P., and Jenness, T., 1998,
{\it Nature}, {\bf 394}, 241.\\
Hu, E., Cowie, L., and McMahon, R., 1998, {\it Astrophysical Journal},
{\bf 502}, L99.\\
Kennicutt, R.C., Keel, W.C., van der Hulst, J.M., Hummel, E., Roettiger,
K.A., 1987, {\em Astronomical Journal}, {\bf 93}, 1011.\\
Koo, D., and Kron, R., 1992, {\em Annu. Rev. Astron. Astrophys.}, {\bf
30}, 613.\\
Larson, R.B., 1974, {\em Mon. Not. R. astr. Soc.}, {\bf 145}, 405.\\
Lilly, S.J., Le F\a'{e}vre, O., Hammer, F., and Crampton, D., 1996, {\it
Astrophysical Journal}, {\bf 455}, 108.\\
Madau, P., Ferguson, H.C., Dickinson, M.E., Giavalisco, M., Steidel,
C.C., Frchter, A., 1996, {\it Mon. Not. R. astr. Soc.}, {\bf 283}, 1388.\\
Madau, P., Pozzetti, L., and Dickinson, M., 1998, {\em Astrophysical
Journal}, {\bf 498}, 106.\\
McCarthy, P., Spinrad, H., Djorgovski, S., Strauss, M. A., van Breugel,
W., and Leibert, J., 1987, {\em Astrophysical Journal}, {\bf 319}, L39.\\
Pahre, M.A., and Djorgovski, S.G., 1995, {\em Astrophysical Journal},
{\bf 449}, L1.\\
Parkes, I. M., Collins, C.A., and Joseph, R.D., 1994, {\em Mon. Not. R.
astr. Soc.}, {\bf 266}, 983.\\
Partridge, R.B., 1974, {\em Astrophysical Journal}, {\bf 192}. 241.\\ 
Partridge, R.B., and Peebles, P.J.E., 1967, {\em Astrophysical Journal}, 
{\bf 147}, 868.\\
Penzias, A.A., and Wilson, R.W., 1965, {\it Atrophysical Journal}, {\bf
142}, 419.\\
Pettini, M., Steidel, C.C., Adelberger, K., Kellogg, M., Dickinson, M.,
and Giavalisco, M., 1998, In {\it Cosmic Origins: Evolution of galaxies,
stars, planets and life}, eds. Woodward, C.E., and Thronson, H.A.,
(Astronomical Society of the Pacific Conference Series), in press.\\
Pritchet, C.J., 1994, {\it Publications of the Astronomical Society of
the Pacific}, {\bf 106}, 1052.\\
Pritchet, C.J., and Hartwick, F.D.A., 1987, {\it Astrophysical Journal},
{\bf 320}, 464.\\ 
Puget, J.L., Abergel, A., Bernard, J.P., Boulanger, F., Burton, W.B.,
D\a'{e}sert, F.X., and Hartmann, D., 1996, {\it Astronomy and Astrophysics},
{\bf 308}, L5.\\
Sandage, A.R., 1995, In {\it The Deep Universe}, eds. Binggeli, B.,
Buser, R., (Springer: Berlin), 1.\\
Smail, I., Ivison, R.J., and Blain, A.W., 1997, {\it Astrophysical
Journal}, {\bf 490}, L5.\\
Smoot, G., and Keay, D., 1993, {\it Wrinkles in time}, (William Morrow:
New York).\\
Steidel, C.C., Giavalisco, M., Pettini, M., Dickinson, M., and
Adelberger, K.L., 1996, {\em Astrophysical Journal}, {\bf 462}, 17.\\
Thompson, D., and Djorgovski, S., 1995, {\it Astronomical Journal}, {\bf
110}, 3.\\
Thompson, D., Djorgovski, S., and Beckwith, S.V.W., 1994, {\em
Astronomical Journal}, {\bf 107}, 1.\\ 
Uson, J.M., Bagri, D.S., and Cornwell, T.J., 1991, {\em Astrophysical
Journal}, {\bf 377}, L65.\\

\newpage

{\bf Biographical Details}

Dr Collins carried out his PhD work at Imperial College, London
University, under the supervision of Dr R. Joseph from 1982-1985. His thesis 
work was based largely on an infrared search for primeval galaxies. Since then 
he has been a research fellow at the University of Edinburgh and a senior
research fellow at the Royal Observatory Edinburgh. In 1992 he moved to
Durham University on a PPARC Advanced Fellowship and in 1994 became a
lecturer at Liverpool John Moores University. He was promoted to Professor in
`Cosmology' in September 1998 . His other research
interests are large-scale structure in the Universe, the evolution of 
galaxies, and the evolution of X-ray emitting galaxy clusters.

\end{document}